\documentclass[floatfix,reprint,amsmath,amssymb,aps,pra]{revtex4-2}
\usepackage{times,mathptmx}
\usepackage{graphicx}
\usepackage[english]{babel}
\usepackage{epstopdf}
\usepackage{longtable}
\usepackage{xcolor}

\begin{document}

\title{Virtual perfect absorption in resonant media and their $\cal PT$-symmetric generalizations}
\author{Denis~V.~Novitsky}
\affiliation{B.~I.~Stepanov Institute of Physics, National Academy of Sciences of Belarus, Nezavisimosti Avenue 68, 220012 Minsk, Belarus}
\email{dvnovitsky@gmail.com}
\author{Alexander~S.~Shalin}
\affiliation{Center for Photonics and 2D Materials, Moscow Institute of Physics and Technology, 141700 Dolgoprudny, Russia \\
Nukus Innovation Institute, I. Karimov 154, 1735401554 Nukus, Karakalpakstan, Uzbekistan \\
Faculty of Physics, M.V. Lomonosov Moscow State University, 119991 Moscow, Russia}

\date{\today}

\begin{abstract}
Virtual perfect absorption (VPA) is an effect simulating real absorption of light by using excitation at a complex frequency corresponding to a scattering zero. We theoretically study VPA in resonantly absorbing and amplifying media irradiated by two counterpropagating waves with exponentially growing amplitudes. We show that VPA critically depends on the medium density (i.e., level of loss or gain) deteriorating in the high-density limit. In contrast, almost ideal VPA persists in the $\cal PT$-symmetric loss-gain bilayer. For high enough gain, the powerful quasilasing pulses are observed at later times symmetrically (single amplifying layer) or asymmetrically ($\cal PT$-symmetric structure) generated in both propagation directions.
\end{abstract}

\maketitle

\section{Introduction}

There is a recent surge of interest in complex-frequency excitation of optical media and nanophotonic structures. Complex frequency $\omega=\omega' +i \omega''$ implies that, besides the usual oscillations of light field governed by $\omega'$, there is a temporal modulation of amplitude due to $\omega''$. For a monochromatic wave, this modulation results in exponentially increasing or decreasing envelope depending on the sign of $\omega''$. Using complex-frequency excitation, one can overcome the difficulties in reaching specific resonances or singularities in system's response. Indeed, the anomalies in light scattering \cite{Krasnok2019} such as poles and zeros are usually associated with complex frequencies and, hence, cannot be excited with standard constant-amplitude (real-frequency) waves. However, the waves with time-varying amplitudes give the possibility to emulate those anomalies extending the frequency space to the entire complex plane.

The first effect based on complex-frequency excitation to be considered in the literature was the virtual perfect absorption (VPA) \cite{Baranov2017}. It is observed under irradiation of a system by the exponentially growing waves, if the rate of exponential growth $\omega''$ is equal exactly to the imaginary part of the scattering zero frequency. This latter condition results in perfect storage of radiation inside the system until the wave's amplitude keeps growing. After the incident radiation is switched off, the stored energy is released to the surrounding space. This is the main difference of VPA from its real-frequency counterpart -- the coherent perfect absorption (CPA). The simplest scheme for VPA observation is a layer irradiated by two counterpropagating exponentially growing waves, but it can be observed for other scatterer geometries such as dielectric cylinders \cite{Baranov2017}, coupled waveguides \cite{Longhi2018}, waveguide-coupled microring resonators \cite{Zhong2020}, and open cavities bounded by properly constructed metasurfaces \cite{Marini2020, Marini2021}. VPA was experimentally realized for mechanical waves \cite{Trainiti2019}, microwave circuits \cite{Marini2022}, and even for ignition of plasma discharge in microwave resonators \cite{Delage2023}. VPA can be used to provide the perfect coupling of incident radiation to the electromagnetic cavity eliminating parasitic reflections. This is the virtual critical coupling analyzed for both the high-quality optical cavities and equivalent microwave transmission lines \cite{Radi2020} as well as for plasma-containing microwave resonators \cite{Delage2022}. As a counterpart to VPA, virtual gain was proposed by harnessing complex-frequency scattering pole with the exponentially decaying incident radiation \cite{Li2020}. It proved to be useful to mimic gain without dealing with real amplifying media, in particular to simulate loss compensation \cite{Kim2023, Guan2023}, parity-time ($\cal PT$) symmetry in the structures with balanced loss and gain \cite{Li2020}, exceptional points and CPA \cite{Farhi2022}, and even the non-Hermitian skin effect in topologically nontrivial lattices \cite{Gu2022}. Excitation at complex frequencies was reported as a means for virtual analogue of pulling optical force \cite{Lepeshov2020} and for light scattering engineering \cite{Kim2022, Ali2021, Loulas2023, Bradley2022}.

In this paper, we extend VPA studies to the field of light interaction with resonant media. We numerically solve the Maxwell-Bloch equations for a layer of two-level medium irradiated by two exponentially growing waves. This work can be considered as a continuation of our recent paper \cite{Novitsky2023} devoted to virtual gain in resonant media. In that case, we have observed the inverted pattern of light distribution under illumination with a single exponentially decaying wave. In contrast, two counterpropagating exponentially growing waves are applied here to observe the VPA effect, when radiation gets stored inside the medium. We show that the nonlinear effects here play a minor role (as opposed to the virtual gain case) with the output radiation profile weakly depending on the incident wave intensity. The efficiency of radiation storage is turned out to be strongly dependent on the medium density (concentration of two-level atoms), so that the VPA gets spoiled for high enough density as shown for both absorbing and amplifying resonant media. The VPA can be restored and observed even in the high-density limit in the $\cal PT$-symmetric structure consisting of two layers with balanced loss and gain. We also discuss the transition to quasilasing regime under exponentially growing excitation in resonant media and $\cal PT$-symmetric bilayers with high enough gain.

\section{Governing equations}

To describe light propagation through the two-level resonant medium, we employ the well-known semiclassical Maxwell-Bloch equations for the atomic polarization $\rho_{12}$, the population difference between the excited and ground levels (inversion) $w=\rho_{22}-\rho_{11}$, and the electric field $E$ \cite{AllenBook, Kalosha1999}
\begin{eqnarray}
\frac{d \rho_{12}}{d t} &=& i \omega_0 \rho_{12} + i \frac{\mu}{\hbar} E w - \frac{\rho_{12}}{T_2}, \label{polar}
\end{eqnarray}
\begin{eqnarray}
\frac{d w}{d t} &=& -4 \frac{\mu}{\hbar} E \textrm{Im} \rho_{12} - \frac{w-w_{eq}}{T_1}, \label{invers}
\end{eqnarray}
\begin{eqnarray}
\frac{\partial^2 E}{\partial z^2}&-&\frac{n^2}{c^2} \frac{\partial^2 E}{\partial t^2} = \frac{4 \pi}{c^2} \frac{\partial^2 P}{\partial
t^2}, \label{Max}
\end{eqnarray}
where $P=2 \mu C \textrm{Re} (\rho_{12})$ is the macroscopic polarization, $n$ is the background refractive index, $\omega_0$ is the resonance frequency, $\mu$ is the transition dipole moment, $T_1$ and $T_2$ are the relaxation times, $C$ is the concentration of two-level atoms, $w_{eq}$ is the equilibrium population difference, $c$ is the speed of light, and $\hbar$ is the Planck constant. The Maxwell-Bloch equations (\ref{polar})--(\ref{Max}) will be numerically solved using essentially the same approach as described in Refs. \cite{Novitsky2009, Novitsky2012}. Further, we adopt the following parameters of the medium: the thickness $L=7$ $\mu$m, the refractive index $n=3$, the relaxation rates $T_1=1$ ns and $T_2=1$ ps, and the medium density parameter $g=4 \pi \mu^2 C/3 \hbar < 0.1$ ps$^{-1}$ corresponding to the concentrations $C \lesssim 2 \cdot 10^{17}$ cm$^{-3}$ at $\mu \sim 10$ D. These values can be reached in the resonant media based on semiconductor quantum dots. Further, we also assume that the carrier frequency is tuned exactly to the resonance with the quantum transition, i.e., $\omega_s=2 \pi c/\lambda_s=\omega_0$.

\begin{figure}[!tb]
\includegraphics[scale=0.5, clip=]{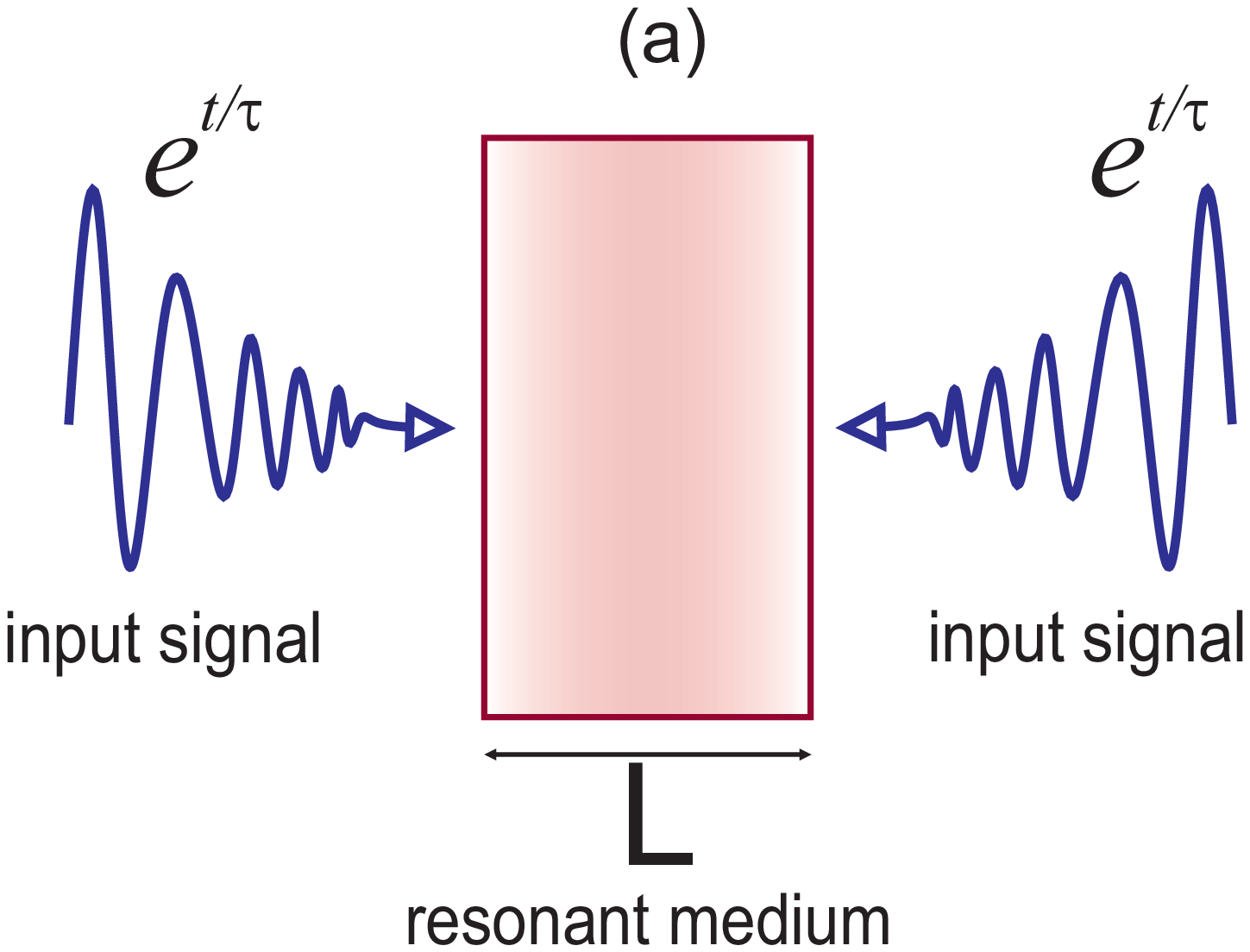}
\includegraphics[scale=1., clip=]{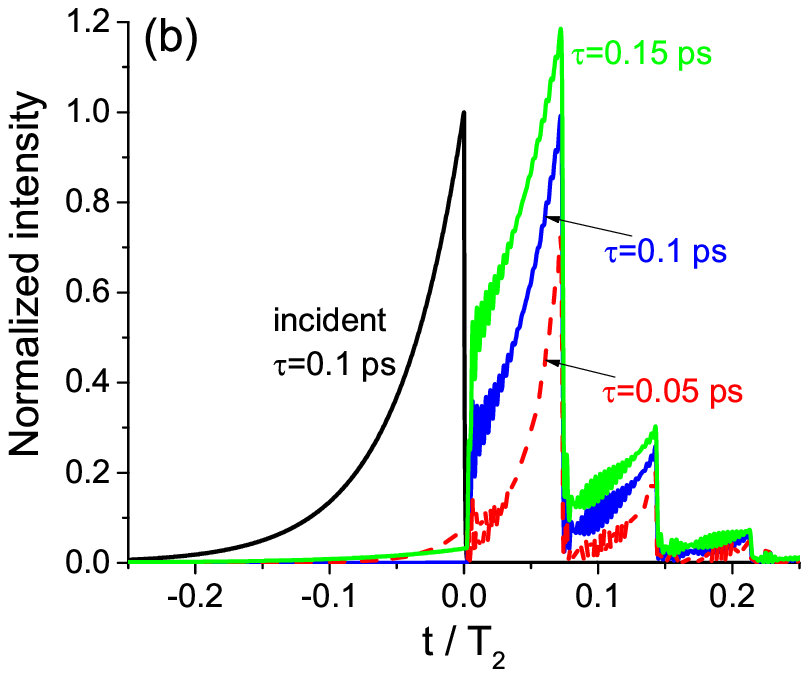}
\caption{\label{fig1} (a) Scheme of two counterpropagating exponentially growing waves incident on the layer of resonant medium. (b) Profiles of radiation intensity (incident and output from the resonantly absorbing medium) at different values of the growth time $\tau$. The output profiles are the same in both (forward and backward) directions.}
\end{figure}

The parameters of incident radiation should be chosen so as to satisfy the conditions for VPA. Following Ref. \cite{Baranov2017}, we consider the two counterpropagating incident waves exponentially growing at first and then rapidly switched off, so that the incident field at the boundary of the medium is
\begin{eqnarray}
E \sim A \left[ e^{t/\tau} \theta(-t) + e^{-t/\tau'} \theta(t) \right], \label{ampl}
\end{eqnarray}
where $\theta(t)$ is the Heaviside step function, $\tau$ is the growth time, and $\tau' \ll \tau$ is the decay time. The scheme of medium excitation with two counterpropagating waves [see Fig. \ref{fig1}(a)] is fundamentally different from the case of virtual gain excited with a single wave \cite{Novitsky2023}. For VPA to occur, the growth rate in Eq. \eqref{ampl} should match the imaginary part of the scattering zero, $1/\tau=\omega''_z$, where
\begin{eqnarray}
\frac{\omega_z}{c} L = \frac{1}{n} \left[ \pi l - i \textrm{ln} \frac{n-1}{n+1} \right]. \label{zero}
\end{eqnarray}
The carrier frequency of the wave \eqref{ampl} is tuned to coincide with the real part of the zero frequency, $\omega_s=\omega'_z=2 \pi c/\lambda_z$, which can be regulated with an integer $l$. For the medium parameters mentioned above and $l=42$, we have $\lambda_z=1$ $\mu$m, whereas $\omega''_z \approx 10$ ps$^{-1}$. Note that these estimates are justified for rather low $g$'s, so that the medium can be approximately treated with the real-valued refractive index $n$.

\section{\label{sec_res}Resonantly absorbing medium}

We start with the case of resonantly absorbing medium with initially all the atoms in the ground state, $w(0)=w_{eq}=-1$. The concentration of atoms is rather low ($g=10^{-4}$ ps$^{-1}$) at first, so that the medium is weakly absorbing and is close to the passive one with the real-valued $n$. Figure \ref{fig1}(b) shows the results of calculations for the counterpropagating incident waves \eqref{ampl} with $\tau'=\tau/100$ [the geometry is shown in Fig. \ref{fig1}(a)]. The intensity dynamics for such exponentially growing waves turn out to be very weakly dependent on its amplitude $A$, which is assumed to be comparatively low corresponding to the maximal Rabi frequency, $\Omega_{\textrm{max}}=\mu A/\hbar = 10^{-4}/T_2$, guaranteeing the absence of medium saturation (most two-level particles remain in the ground state). One can readily see that the case $\tau=1/\omega''_z=0.1$ ps clearly matches the condition for VPA: while the incident amplitude is exponentially growing, there is no transmission and reflection, so that the radiation is stored inside the medium; only after radiation is turned off, the stored energy rapidly leaves the layer in the form of short bunches of light going symmetrically in both directions due to symmetric excitation with the two counterpropagating waves (only forward radiation is shown in Fig. \ref{fig1}, the same curves are obtained for the backward radiation). For slower or faster growing amplitudes, radiation is partly transmitted through the layer even before switching-off moment. Nevertheless, in these cases, the sharp pulses are also formed after switching incident radiation off with the same duration and the peak intensity clearly dependent on $\tau$: it is higher for longer $\tau$, since such pulses contain more energy.

\begin{figure}[!tb]
\includegraphics[scale=0.9, clip=]{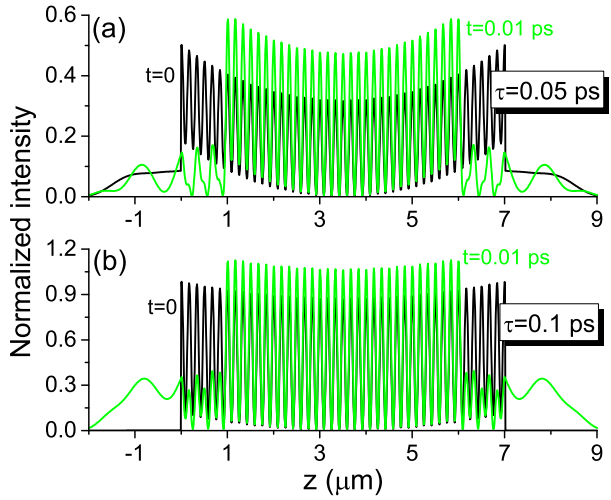}
\caption{\label{fig2} Spatial distributions of radiation at different time instants -- at the switching-off moment ($t=0$, black line) and a little later [$t=0.01$ ps, green (grey) line]. The layer covers the region $0<z<7$ $\mu$m.}
\end{figure}

Storage and subsequent release of radiation can be assessed in Fig. \ref{fig2}, where the spatial localization is shown at different instants of time. At the switching-off moment ($t=0$), radiation is mostly localized inside the layer of resonant medium, this localization being almost perfect for $\tau=0.1$ ps [Fig. \ref{fig2}(b)]. The number of intensity peaks inside the layer is given by the zero order $l$. This localization is far from perfect for shorter $\tau$ demonstrating significant intensity outside the layer and the parabolic envelope inside it with higher intensity near the edges of the layer [Fig. \ref{fig2}(a)]. After the signal is switched off, radiation rapidly leaves the layer in both directions. It is interesting that this also results in some compression of the intensity comb closer to the layer center with the corresponding increase of intensity. This is obviously due to radiation propagating not only outside, but also inside the layer. Subsequent release of this compressed radiation causes the secondary bursts of output radiation seen in Fig. \ref{fig1}.

\begin{figure}[!tb]
\includegraphics[scale=0.95, clip=]{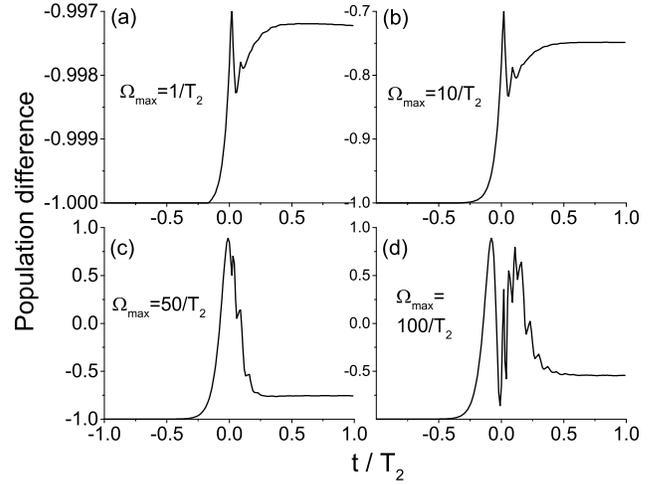}
\caption{\label{fig3} Dynamics of population difference inside the absorbing medium (the point at $z=L/4$ from the entrance) for the different values of $\Omega_{\textrm{max}}$. The growth rate $\tau=0.1$ ps.}
\end{figure}

So far, our discussion has not touched any peculiarities characteristic for the resonant media. The same behaviour of exponentially growing fields is generally expected for passive media as well. To find out whether the resonant medium is influenced, we study the dynamics of population difference $w$ for ever larger peak amplitude $\Omega_{\textrm{max}}$ as shown in Fig. \ref{fig3}. It can be seen that under the action of radiation, the medium absorbs some radiation and passes from the ground state ($w=-1$) to a partially excited state. Excitation is more efficient for larger $\Omega_{\textrm{max}}$ and follows the exponential envelope of the incident waveform. After the incident signal is turned off, a significant part of the absorbed radiation is re-emitted and leaves the medium, and the population difference decreases, but never reaches the original value. The specific dynamics of population difference strongly depend on the radiation intensity. For the relatively low intensities, the dynamics are essentially the same (if the scale is corrected), as shown in Figs. \ref{fig3}(a) and \ref{fig3}(b). If the intensity is high enough to almost invert the medium, the dynamics reminiscent of dull excitation-deexcitation cycle can be observed with the notches clearly connected to the bunches of output radiation [Fig. \ref{fig3}(c)]. For even higher intensities, the dynamics can be very complicated as shown in Fig. \ref{fig3}(d). It should be noted that all these differences in the behavior of $w$ have practically no effect on the radiation profile at the layer exit because of low concentration of absorbing particles. The population difference can also be regulated with the rate of exponential growth, since for shorter $\tau$, radiation is less monochromatic and the resonant medium will be less excited.

\begin{figure}[!tb]
\includegraphics[scale=1., clip=]{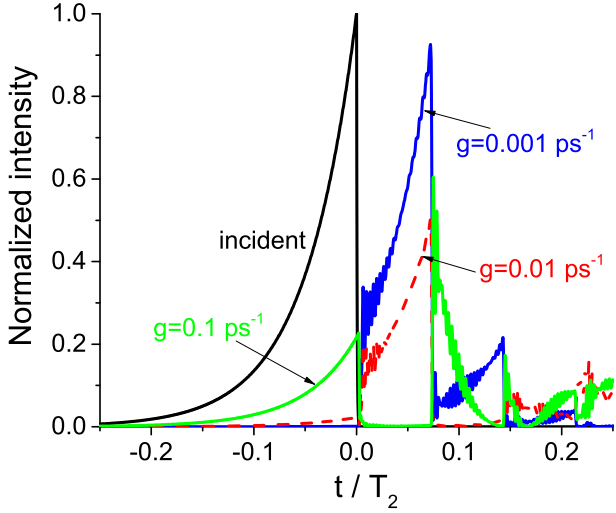}
\caption{\label{fig4} Profiles of radiation intensity at different values of the density parameter $g$. The exponential growth time is $\tau=0.1$ ps, the amplitude is $\Omega_{\textrm{max}}=10^{-4}/T_2$.}
\end{figure}

Let us now consider the influence of absorption on VPA in resonant media. To that end, we increase the medium density parameter $g$ and recalculate interaction of exponentially growing radiation ($\tau=0.1$ ps) with the medium. The results shown in Fig. \ref{fig4} indicate that VPA remains virtually intact for $g=10^{-3}$ ps$^{-1}$, except for the small decrease of the output peak intensity. For larger values of $g$, absorption clearly violates the VPA resulting in the appearance of output radiation at $t<0$ and the overall decrease of peak intensities. The calculations show that for a given $g$, there is the value of $\tau$ providing minimal deviation from VPA. For example, for $g=10^{-2}$ ps$^{-1}$, such minimum is reached at $\tau_{\textrm{min}}=0.18$ ps. One could try to estimate this value using Eq. \eqref{zero} with the refractive index $n_c$ corrected to take absorption into account. The stationary approximation gives for such a corrected permittivity the expression as follows: $\varepsilon=n^2_{c} \approx n^2+3 i l^2 g T_2$ \cite{Novitsky2017}. However, such an estimate does not give a correct value of $\tau_{\textrm{min}}$, because the exponentially growing waveforms are evidently far from the steady-state excitation, so that the very notion of refractive index looses its meaning. 

\begin{figure}[!tb]
\includegraphics[scale=1., clip=]{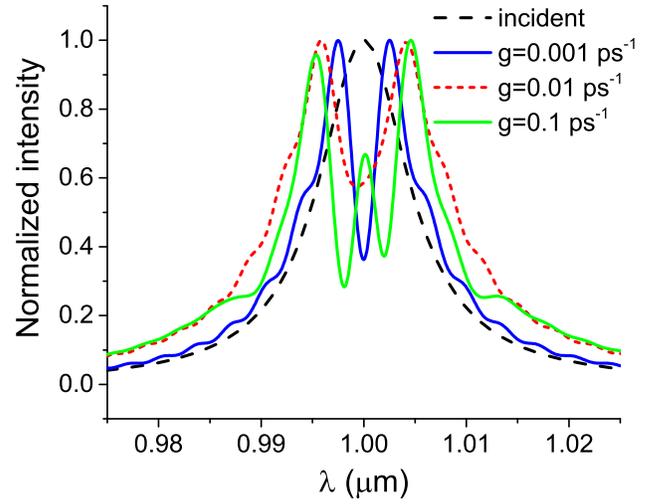}
\caption{\label{fig5} Spectra of incident and transmitted radiation at different values of the density parameter $g$.}
\end{figure}

Another interesting feature seen in Fig. \ref{fig4} is that for large densities ($g=0.1$ ps$^{-1}$), the intensity profile differs rather sharply from those observed for more diluted media. In particular, there is a time interval right after the switching-off moment with the output radiation almost absent. Such a change indicates a different regime of light-matter interaction that can be corroborated with the spectra of output radiation shown in Fig. \ref{fig5}. The spectrum of initial radiation can be analytically estimated with the Fourier transform of Eq. \eqref{ampl}:
\begin{eqnarray}
I(\omega)&=&\left| \int_{-\infty}^\infty E(t) e^{-i \omega t} dt \right|^2 \sim \nonumber \\
&&\sim \left| \frac{1}{1/\tau-i(\omega-\omega_s)} + \frac{1}{1/\tau'+i(\omega-\omega_s)} \right|^2 \approx \nonumber \\
&&\approx \frac{1}{1/\tau^2+(\omega-\omega_s)^2}. \label{spectr}
\end{eqnarray}
For $g=10^{-4}$ ps$^{-1}$, the output spectrum essentially repeats the spectrum of initial radiation. For larger values of $g$, the spectrum gets wider and the dip is clearly seen at the resonant wavelength ($\lambda=1$ $\mu$m) indicating strong absorption. Finally, for $g=0.1$ ps$^{-1}$, the dip splits as characteristic for the strong light-matter coupling. Thus, one obtains rather different types of intensity profile depending on the regime governed by the two-level atoms density.

We should note that the discontinuous envelope governed by Eq. \eqref{ampl} with the very fast decay ($\tau' \ll \tau$) is not necessary to observe the main features of the VPA as shown in the Appendix. Nevertheless, we use such a description to obtain this effect in the most clear and sharp way.

\section{Resonantly amplifying medium}

\begin{figure}[!tb]
\centering
\includegraphics[scale=1., clip=]{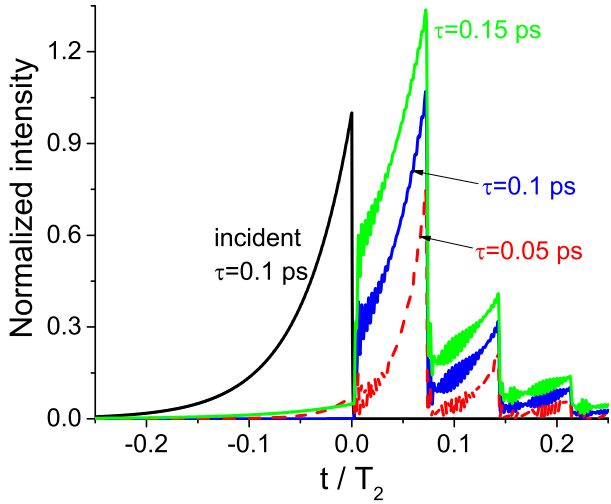}
\caption{\label{fig6} Profiles of radiation intensity (incident and output from the resonantly amplifying medium) at the different values of growth time $\tau$. The output profiles are the same in both (forward and backward) directions.}
\end{figure}

\begin{figure}[!tb]
\includegraphics[scale=0.95, clip=]{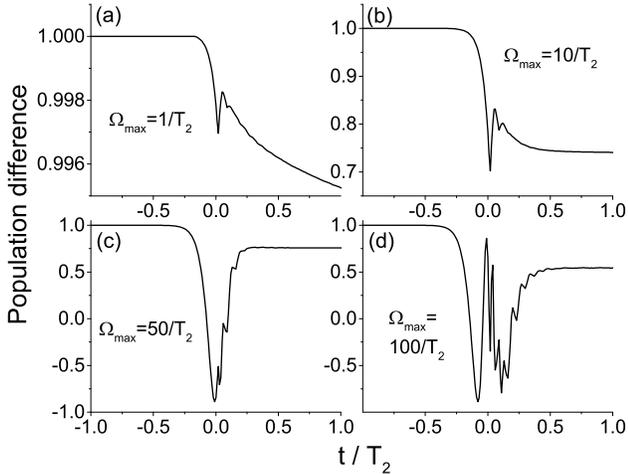}
\caption{\label{fig7} Dynamics of population difference inside the amplifying medium (the point at $z=L/4$ from the entrance) for the different values of $\Omega_{\textrm{max}}$. The growth rate $\tau=0.1$ ps.}
\end{figure}

In this Section, we consider the medium initially fully inverted, $w(0)=w_{eq}=1$, when all the two-level atoms are at the excited level. Again, we start with the small radiation amplitude $\Omega_{\textrm{max}}= 10^{-4}/T_2$ and the relatively low medium density ($g=10^{-3}$ ps$^{-1}$) to guarantee weak gain. The results of output profiles for different growth times $\tau$ are shown in Fig. \ref{fig6}. It is seen that $\tau=0.1$ ps still corresponds to VPA with no radiation before the exponential growth is switched off and slightly amplified burst of radiation afterwards (compare with Fig. \ref{fig1}). Similar amplification can be noticed for other values of $\tau$ as well. For larger amplitudes of incident waveform, the relative amplification gets weaker becoming less appreciable against the background of a strong signal, so that the output profiles tend to be closer to those in Fig. \ref{fig1}. At the same time, the population difference strongly deviates from the initial value as we increase $\Omega_{\textrm{max}}$ as shown in Fig. \ref{fig7}. The resulting dynamics of $w$ are essentially inverse of those obtained in Fig. \ref{fig3} for the absorbing medium.

\begin{figure}[!tb]
\includegraphics[scale=1., clip=]{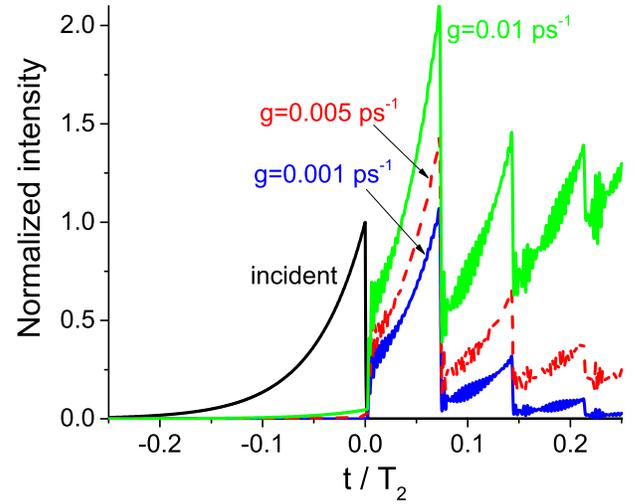}
\caption{\label{fig8} Profiles of radiation intensity at different values of the density parameter $g$. The exponential growth time is $\tau=0.1$ ps, the amplitude is $\Omega_{\textrm{max}}=10^{-4}/T_2$.}
\end{figure}

\begin{figure}[!tb]
\includegraphics[scale=0.95, clip=]{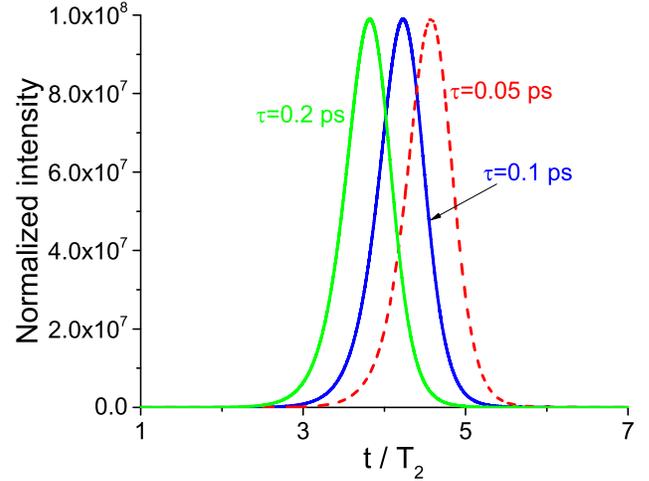}
\caption{\label{fig9} Profiles of quasilasing radiation intensity at $g=10^{-2}$ ps$^{-1}$ for different values of growth time $\tau$. Note the increased scale along the vertical axis.}
\end{figure}

Increase in density parameter $g$ allows us to observe higher gain as shown in Fig. \ref{fig8}. Similar to the results in the previous Section, VPA is clearly violated at larger $g$'s. Moreover, for large enough density parameters, the output intensity continues to grow that is especially noticeable for $g=10^{-2}$ ps$^{-1}$. This marks the large-gain regime, which we call here quasilasing in line with the observations of Ref. \cite{Novitsky2017}. In this regime, the energy pumped in the medium is released in the form of strong burst of radiation. In contrast to usual lasing, in our case, the trigger for energy release is not the noise seed, but the incident radiation of relatively strong intensity. As shown in Fig. \ref{fig9}, the form and energy content of quasilasing pulses are the same for different growth times, but the peak position can be clearly shifted: The longer the $\tau$, the sooner the pulse appears at the output.

\section{${\cal PT}$-symmetric bilayer}

In the previous sections, we dealt with VPA in the single layers of absorbing and amplifying resonant media. Here, we divide the layer into two sublayers -- one with loss and another with gain, so that the total thickness of the resulting bilayer remains intact. We assume that the equilibrium population difference has the same magnitude $|w_{eq}|$ in both sublayers and differs only in sign (negative in the loss layer and positive in the gain one). The balance between loss and gain guarantees that the bilayer obeys ${\cal PT}$ symmetry.

\begin{figure}[!tb]
\includegraphics[scale=1., clip=]{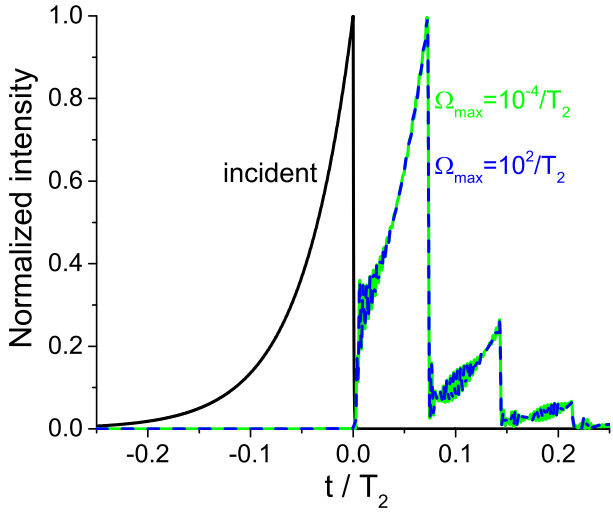}
\caption{\label{fig10} Profiles of radiation intensity transmitted through the ${\cal PT}$-symmetric bilayer are essentially the same at different values of the amplitude $\Omega_{\textrm{max}}$. The exponential growth time is $\tau=0.1$ ps, the equilibrium population difference is $|w_{eq}|=1$.}
\end{figure}

\begin{figure}[!tb]
\includegraphics[scale=0.95, clip=]{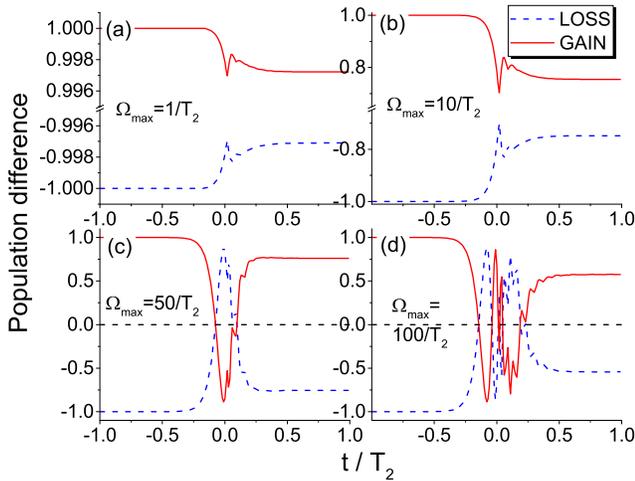}
\caption{\label{fig11} Dynamics of population difference in the middle of loss and gain layers for the different values of $\Omega_{\textrm{max}}$. The growth time $\tau=0.1$ ps, the equilibrium population difference is $|w_{eq}|=1$.}
\end{figure}

As previously, we start with the case of weak signal $\Omega_{\textrm{max}}=10^{-4}/T_2$ and low density parameter $g=10^{-4}$ ps$^{-1}$ guaranteeing small loss and gain even for $|w_{eq}|=1$. The population difference is not changing under these conditions and the structure remains ${\cal PT}$-symmetric as before the incident radiation comes. As a result, for the exponentially growing counterpropagating waves with $\tau=0.1$ ps, we obtain VPA as in the passive medium. If we now increase the amplitude of waves, the resulting output profiles still demonstrate perfect VPA as shown in Fig. \ref{fig10} and, in fact, are independent of the value of $|w_{eq}|$. However, the population difference can significantly change under strong incident radiation as demonstrated in Fig. \ref{fig11} (compare also with Figs. \ref{fig3} and \ref{fig7} for single loss and gain layers). It is seen that although the dynamics of $w$ can be quite complex, its change remain consistent in loss and gain layers, so that the bilayer can be considered as ${\cal PT}$-symmetric at every time instant.

\begin{figure}[!tb]
\includegraphics[scale=1., clip=]{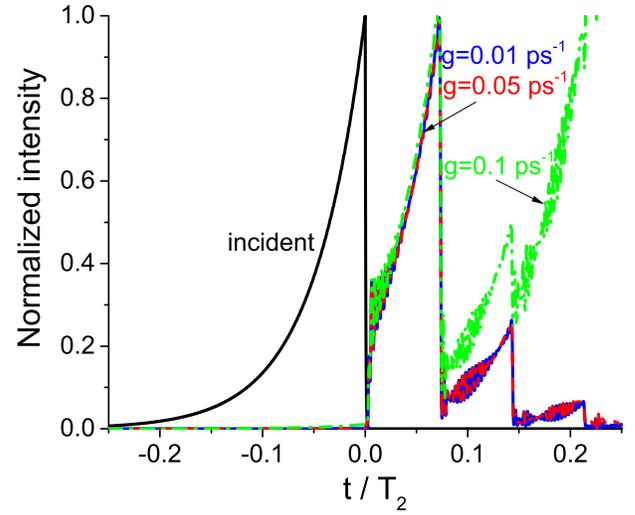}
\caption{\label{fig12} Profiles of radiation intensity transmitted through the ${\cal PT}$-symmetric bilayer at different values of the density parameter $g$. The exponential growth time is $\tau=0.1$ ps, the amplitude is $\Omega_{\textrm{max}}=10^{-4}/T_2$, the equilibrium population difference is $|w_{eq}|=1$.}
\end{figure}

Let us now increase the density parameter $g$, so that the loss and gain increase as well. The results for different values of $g$ are shown in Fig. \ref{fig12}. In contrast to the calculations for the single absorbing and amplifying layers reported in Figs. \ref{fig4} and \ref{fig8} respectively, we see almost perfect VPA up to $g=0.1$ ps$^{-1}$ with no output radiation before the input is switched off. Moreover, the first burst of output radiation has essentially the same profile and peak intensity for any $g$. The reason is obviously connected to the balanced loss and gain making the ${\cal PT}$-symmetric structure similar to the passive layer without any loss or gain at all. Thus, using such systems with compensated loss and gain allows one to widen the VPA conditions to higher densities of resonant atoms.

\begin{figure}[!tb]
\includegraphics[scale=1., clip=]{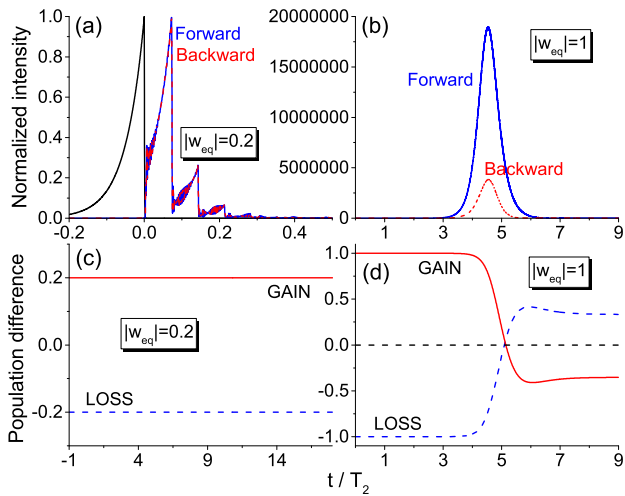}
\caption{\label{fig13} (a) and (b) Profiles of radiation intensity transmitted through the ${\cal PT}$-symmetric bilayer at different values of the equilibrium population difference $|w_{eq}|$: the same forward and backward profiles at $|w_{eq}|=0.2$ and significanly different profiles at $|w_{eq}|=1$. (c) and (d) Corresponding dynamics of population difference in the middle of loss and gain layers. The exponential growth time is $\tau=0.1$ ps, the amplitude is $\Omega_{\textrm{max}}=10^{-4}/T_2$, the density parameter is $g=0.05$ ps$^{-1}$.}
\end{figure}

One can note the growth in intensity at larger times (see Fig. \ref{fig12} at $g=0.1$ ps$^{-1}$) indicating transition to the quasilasing regime as we have seen for the single amplifying layer. This transition is associated with ${\cal PT}$-symmetry breaking as we show further. In Fig. \ref{fig13}, we perform the calculations for different values of $|w_{eq}|$ governing the value of imaginary part of permittivity. The density parameter is took to be large enough, $g=0.05$ ps$^{-1}$. One can see that standard VPA is observed at $|w_{eq}|=0.2$ [Fig. \ref{fig13}(a)], the corresponding population difference being constant [Fig. \ref{fig13}(c)]. In contrast, the powerful pulses are generated inside the structure at $|w_{eq}|=1$ [Fig. \ref{fig13}(b)], whereas the population difference strongly changes [Fig. \ref{fig13}(d)]. It is interesting that the pulses are generated asymmetrically in the forward and backward directions. This is due to asymmetry of the structure itself, which consists of loss and gain layers. If the layers are swapped (i.e., we have the gain layer first and then the loss layer), the forward and backward pulses are also interchanged. This means that the propagation direction locking reported previously for ${\cal PT}$-symmetric multilayers \cite{Novitsky2018} is absent in our case, because the structure is symmetrically excited from both sides. The asymmetry is seen not only in the output radiation, but also in the population dynamics [Fig. \ref{fig13}(d)], so that the final values of $w$ in the layers have different magnitudes being approximately $0.332$ and $-0.353$, respectively (in contrast to the perfect correspondence in Fig. \ref{fig11}). Note also that the result of this dynamics is that loss and gain layers effectively switch their places: the gain layer looses its inversion radiating energy, whereas the loss layer absorbs part of this energy pushing $w$ above zero.

\begin{figure}[!tb]
\includegraphics[scale=0.9, clip=]{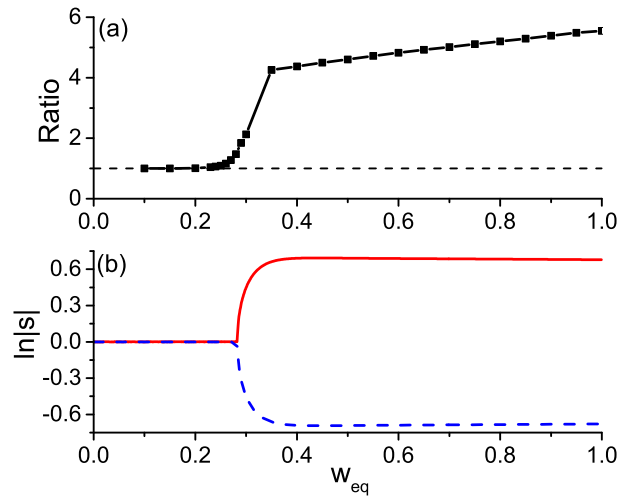}
\caption{\label{fig14} (a) Ratio of radiation energy going out the ${\cal PT}$-symmetric bilayer in the forward and backward directions as a function of the equilibrium population difference $|w_{eq}|$. (b) Behavior of logarithm of the scattering-matrix eigenvalues.}
\end{figure}

The connection of the quasilasing regime to ${\cal PT}$-symmetry breaking can be proved by considering the eigenvalues of the scattering matrix. We use the scattering matrix in the form as follows: $\hat S_1 = \left( \begin{array}{cc} r_F & t \\ t & r_B \end{array} \right)$. This definition is convenient for our aim, since the exceptional points in this case can be associated with the lasing threshold \cite{Novitsky2020}. Here, $t$ is the transmission coefficient and $r_{F,B}$ are the reflection coefficients for the waves incident in the forward and backward directions, respectively. These coefficients can be calculated with the standard transfer-matrix method taking the permittivities of loss and gain layers as $\varepsilon \approx n^2 \pm 3 i l^2 g T_2 |w_{eq}|$ \cite{Novitsky2017}. The resulting dependence of eigenvalues logarithm on $|w_{eq}|$ is shown in Fig. \ref{fig14}(b). The exceptional point dividing the ${\cal PT}$-symmetric and broken-${\cal PT}$-symmetry phases is clearly seen, where the eigenvalues turn from the unimodular values ($|s_{1,2}|=1$) to the inverse values ($|s_1|=1/|s_2|$) \cite{Ge2012}. On the other hand, in Fig. \ref{fig14}(a), we show the ratio of energies radiated in the forward and backward directions: it is equal to unity for the symmetric output (standard VPA) and larger than unity for the asymmetric generation of powerful pulses (quasilasing regime). Comparing Figs. \ref{fig14}(a) and \ref{fig14}(b), one can easily verify that the transition to the quasilasing regime is associated with ${\cal PT}$-symmetry breaking at the exceptional point ($|w_{eq}| \approx 0.28$). Thus, the usual scattering-matrix approach is applicable to predict the transition from the VPA to the quasilasing regime in ${\cal PT}$-symmetric layered systems.

\section{Conclusion}

In this paper, the concept of VPA observed for two counterpropagating waves with the complex frequency at the scattering zero is applied to the resonant two-level media -- either absorbing or amplifying. We show that VPA is preserved in such nonlinear media, at least for low enough concentrations of two-level particles. In the latter case, VPA is almost independent of light intensity, although the dynamics of level populations can be strongly perturbed. For higher concentrations corresponding to larger coefficients of loss or gain, some part of radiation leaks out of the medium even before the incident waves are switched off. This violation of VPA can be eliminated in a ${\cal PT}$-symmetric generalization of the medium consisting of two sublayers with balanced loss and gain. In the high-gain limit (above the exceptional point in the ${\cal PT}$-symmetric system), pulsed quasilasing can be also initiated after the incident radiation is switched off. The results reported in the paper will be of interest for applications of strongly localized radiation in nonlinear and laser photonics as well as for studies of non-Hermitian systems.

\acknowledgements{D.V.N. is grateful to Prof. Alex Krasnok for fruitful discussions and suggestions. The authors gratefully acknowledge the financial support from the Belarusian Republican Foundation for Fundamental Research (Project No. F22TURC-001) and the Ministry of Science and Higher Education of the Russian Federation (Agreement No. 075-15-2022-1150). A.S.S. acknowledges the support of
the Latvian Council of Science (Project: NEO-NATE, No. lzp-2022/1-0553). The calculations of non-Hermitian multilayered structures are partially supported by the Russian Science Foundation (Grant No. 23-72-00037).}

\appendix

\section{Influence of the incident waves shape}

\begin{figure}[!tb]
\includegraphics[scale=0.9, clip=]{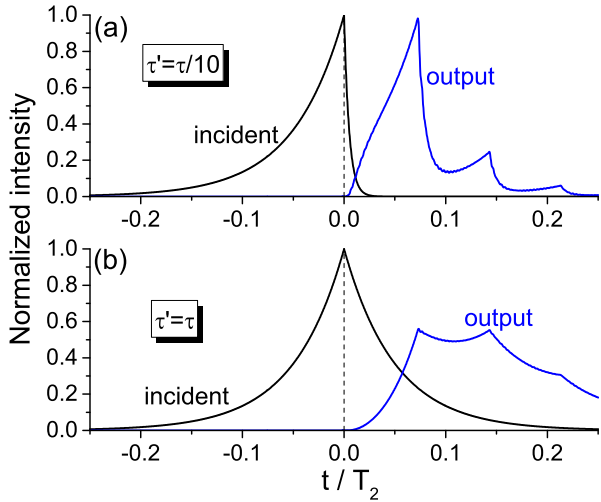}
\caption{\label{fig15}  Profiles of radiation intensity (incident and output from the resonantly absorbing medium) at different values of the decay time (a) $\tau'=\tau/10$ and (b) $\tau'=\tau$ with $\tau=0.1$ ps. Other parameters are the same as in Section \ref{sec_res}.}
\end{figure}

\begin{figure}[!tb]
\includegraphics[scale=0.9, clip=]{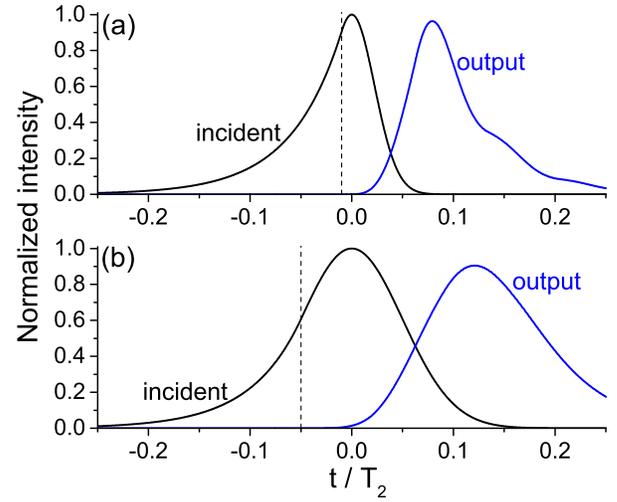}
\caption{\label{fig16}  Profiles of radiation intensity (incident and output from the resonantly absorbing medium) for the smoothed incident field. The vertical dashed line shows the instant of transition from the exponential curve to the Gaussian one. The Gaussian time is (a) $\tau''=0.45\tau$ and (b) $\tau'=\tau$ with $\tau=0.1$ ps. Other parameters are the same as in Section \ref{sec_res}. }
\end{figure}

In the main part of the paper, we have studied the VPA under excitation with the field described by Eq. \eqref{ampl}. The main requirement for the VPA to occur is the exponential growth of the field, so that radiation is released from the medium only after this growth is switched off with the characteristic time $\tau'$. We have assumed that $\tau'$ is much shorter that the growth time $\tau$. Such fast switch-off may be hard to realize in practice. In Fig. \ref{fig15}, we show that essentially the same behavior is observed for $\tau' \sim \tau$ as well. The output radiation appears only after the exponential growth is stopped. The value of $\tau'$ influences the shape and intensity of the output radiation, so that the shorter $\tau'$ the higher output intensity.

The main features of VPA are saved even if we consider not the discontinuous field \eqref{ampl}, but the smoothed one. As an example, Fig. \ref{fig16} shows the calculations for the field consisting of the exponentially growing piece and the Gaussian piece stitched together at the time instant marked with the dashed line. The Gaussian time $\tau''$ appearing in the envelope expression $E \sim A e^{-t^2/2\tau''^2}$ can be derived from the conditions on the continuity of the field and its first derivative and depends on the instant of transition from the exponential curve to the Gaussian one with respect to the instant of peak intensity. One can readily see in Fig. \ref{fig16} that as expected for the VPA, the output field appears only after the exponentially growing field is switched off. The shorter $\tau''$ (hence, the closer the dashed line to the peak of the curve) the closer calculations to those reported in the main text of this paper.

\end{document}